\documentclass[twocolumn]{aastex631}

\usepackage{units}

\usepackage{amsmath}

\newcommand{\code}[1]{\texttt{#1}}
\newcommand{\mesa}{\code{MESA}}
\newcommand{\MESA}{\mesa}

\newcommand{\Msun}{\ensuremath{\mathrm{M}_\odot}}
\newcommand{\gcc}{\ensuremath{\mathrm{g\,cm^{-3}}}} % density units

\newcommand{\Mch}{\ensuremath{\mathrm{M}_{\rm Ch}}}

\newcommand{\Ye}{\ensuremath{Y_{\rm e}}}
\newcommand{\EF}{\ensuremath{E_{\rm F}}}

\newcommand{\Tc}{\ensuremath{T_{\rm c}}}
\newcommand{\logTc}{\ensuremath{\log(T_{\rm c}/\rm K)}}

\newcommand{\Msunyr}{\ensuremath{\Msun\,\mathrm{yr^{-1}}}}

\newcommand{\Dif}{\ensuremath{\mathrm{D}}}

\newcommand{\ddt}[1]{\frac{\partial #1}{\partial t}} %partial time derivative 
\newcommand{\DDt}[1]{\frac{\Dif #1}{\Dif t}} % Lagrangian time derivative
\newcommand{\ddm}[1]{\frac{\partial #1}{\partial m}} %partial derivative wrt m
\newcommand{\ddr}[1]{\frac{\partial #1}{\partial r}} %partial time derivative 

\newcommand{\epsgrav}{\ensuremath{\epsilon_{\mathrm{grav}}}} % Gravitational heating rate
\newcommand{\epsnunuc}{\ensuremath{\epsilon_{\nu, \rm nuc}}} % Neutrino loss rate
\newcommand{\epsnuth}{\ensuremath{\epsilon_{\nu, \rm th}}} % Neutrino loss rate
\newcommand{\epsnuc}{\ensuremath{\epsilon_{\mathrm{nuc}}}} % Neutrino loss rate
\newcommand{\NA}{\ensuremath{N_{\mathrm{A}}}} % Avagadro
\newcommand{\amu}{\ensuremath{m_{\mathrm{u}}}} % amu
\newcommand{\NB}{\ensuremath{N_{\mathrm{B}}}} % Baryon
\newcommand{\nB}{\ensuremath{n_{\mathrm{B}}}} % Baryon

\newcommand{\Zbar}{\ensuremath{\bar{Z}}}

\newcommand{\gradad}{\ensuremath{\nabla_{\rm ad}}}
\newcommand{\gradT}{\ensuremath{\nabla_{T}}}

\newcommand{\Mcc}{\ensuremath{M_{\rm cc}}}
\newcommand{\rcc}{\ensuremath{r_{\rm cc}}}
\newcommand{\recap}{\ensuremath{\lambda_{\rm ec}}}
\newcommand{\rbeta}{\ensuremath{\lambda_{\beta}}}

\newcommand{\nuecap}{\ensuremath{\epsilon_{\nu, \rm ec}}}
\newcommand{\nubeta}{\ensuremath{\epsilon_{\nu, \beta}}}

\newcommand{\nuclei}[2]{\ensuremath{\mathrm{^{#1}#2}}}

\newcommand{\carbon}[1][12]{\nuclei{#1}{C}}
\newcommand{\nitrogen}[1][14]{\nuclei{#1}{N}}
\newcommand{\oxygen}[1][16]{\nuclei{#1}{O}}
\newcommand{\fluorine}[1][19]{\nuclei{#1}{F}}
\newcommand{\neon}[1][20]{\nuclei{#1}{Ne}}
\newcommand{\sodium}[1][23]{\nuclei{#1}{Na}}
\newcommand{\magnesium}[1][24]{\nuclei{#1}{Mg}}

\renewcommand*{\vec}[1]{\boldsymbol{#1}}
\newcommand{\grad}{\vec{\nabla}}
\newcommand{\vecdot}{\vec{\cdot}}
\renewcommand*{\emph}[1]{\textit{\textbf{#1}}}

\newcommand{\mesaone}{Paper~I}  % the first mesa paper
\newcommand{\mesafour}{Paper~IV} % the third mesa paper
\begin{document}

\author[0000-0002-4870-8855]{Josiah Schwab}
\affiliation{Department of Astronomy and Astrophysics, University of California, Santa Cruz, CA 95064, USA}
\altaffiliation{Humble Fellow}
\correspondingauthor{Josiah Schwab}
\email{jwschwab@ucsc.edu}

\title{Some Thoughts on the Convective Urca Process}

\begin{abstract}
  I have repeatedly grappled with the question of
  how the convective Urca process affects stellar evolution, in
  particular during the high-density convective carbon burning that
  can occur in near-Chandrasekhar-mass white dwarfs.  This manuscript
  collects some fragmentary thoughts from various failed and
  abandoned attempts.  This is not a complete work, does not provide
  a comprehensive overview of the literature, and has no definitive
  conclusions.  It is posted in the hope that some part of it might
  prove useful to someone at some point in the future.  I also take
  this opportunity to include anotter important result of more general
  interest (Appendix~\ref{sec:deathotter}).
\end{abstract}


\section{Overview}

The explosion of a (near) Chandrasekhar-mass, carbon-oxygen white dwarf
(\Mch\ CO WD) has long been understood as a potential progenitor of
Type Ia supernovae \citep[see review by][and references
therein]{Hillebrandt2000}.  In this scenario, the WD slowly grows via
accretion of mass from a non-degenerate companion star.  When the mass
of the WD has increased sufficiently, such that its central
temperature and density reach the conditions for carbon ignition, a
phase of convective central carbon burning begins.  This phase, often
known as ``simmering'', eventually ends when the burning becomes
dynamical, leading to the formation of a deflagration and
subsequently the explosion \citep[e.g.,][]{Woosley2004}.

Stellar evolution calculations aim to simulate the progenitor systems
over the long timescales of WD growth up to as close to the moment of
explosion as their modeling assumptions reliably allow.  One of the
long-standing challenges in following models through the simmering
phase is the uncertain effects of the convective Urca process.  An
Urca process is a cyclic series of electron-capture and beta-decay
reactions, each of which emits a neutrino that then free-streams from
the star \citep{Gamow1940, Gamow1941}. The central convection zone
during carbon burning can span several orders of magnitude in density,
such that for some isotopes, electron-capture reactions are favored at
the center and beta-decay reactions near the outer edge.  Convective
mixing enables the transport of material between these regions,
leading to the operation of the convective Urca process
\citep{Paczynski1972d, Paczynski1973a}.  The precise effects of the
convective Urca process are a matter of long debate.  But at a schematic level,
the presence of this additional energy loss mechanism requires
additional carbon to be burned in order to reach explosion.  The total
amount of carbon burned influences the neutron excess of the material
\citep[e.g.,][]{Chamulak2008} and thus the nucleosynthesis during the
explosion.

The most detailed observational probe of Type Ia nucleosynthesis comes
from elemental and isotopic abundance measurements made in supernova
remnants.  Because the nucleosynthesis is sensitive to the mass and
metallicity of the exploding WD, this provides an opportunity to test
and constrain explosion models \citep[e.g.,][]{Badenes2008,MartinezRodriguez2017}.

Theoretical predictions for the detailed nucleosynthesis the
Chandrasekhar-mass Type Ia channel are dependent on an accurate
treatment of the convective Urca process.
A self-consistent treatment of the convective Urca process suitable
for incorporation in stellar evolution codes is sorely needed; this
would unlock the potential of existing and future observations to
constrain supernova progenitor systems.

In this paper, we discuss the potential effects of the convective Urca
process and the challenges of modeling it in stellar evolution codes.

\section{Setup}

We use Modules for Experiments in Stellar Astrophysics
\citep[\MESA;][]{Paxton2011, Paxton2013, Paxton2015, Paxton2018,
  Paxton2019} to construct stellar models of CO WDs during the
simmering phase.  Most of the calculations shown in this paper were
performed in early 2020 with \MESA\ development versions functionally
equivalent to release r12778.  Attempting to model high-density
convective carbon burning with \MESA\ proved to be a challenging
problem that revealed a variety of bugs, inconsistencies, and
shortcomings.  A reader wishing to further explore these types of
models should use a forthcoming \MESA\ release (so one made later
than October 2021) and carefully and skeptically evaluate the 
assumptions and outputs.  In Appendix~\ref{sec:laws}, we provide a
detailed discussion of the key \MESA\ equations.

\subsection{Initial Model}

We use a toy model of an accreting CO WD that reaches central carbon
ignition and continues into convective carbon burning.  We start with
a homogeneous $\unit[1]{\Msun}$ CO WD consisting of 70\% \carbon[12]
and 30\% \oxygen[16] by mass.  This object then accretes material of
the same composition at a rate
$\dot{M} = \unit[3\times10^{-7}]{\Msunyr}$.  During this evolution,
compressional heating and neutrino cooling largely erase its previous
history \citep[e.g.,][]{Paczynski1971d, Brooks2016}, so the choice of
initial mass and central temperature are not particularly important.
In more detailed models, there remains some diversity in the
properties at ignition due to variations in the initial mass and
initial central temperature \citep{Lesaffre2006, Chen2014b}.

\subsection{Input physics}

We use the simple N1 nuclear network from \citet{Forster2010}, which
is specifically tailored for this phase of high-density hydrostatic
carbon burning.  Our nuclear reaction rates are primarily the \MESA\
default rates drawn from JINA REACLIB \citep{Cyburt2010}, modified by
the screening prescription of \citet{Chugunov2007}.  This is not a
state-of-the-art carbon ignition curve \citep[see
e.g.,][]{Gasques2005, Yakovlev2006}, but serves our purposes.
We also use the \citet{Suzuki2016a} reaction rate tables for the Urca
process weak reactions and the \nitrogen[13]$(e^-,\nu_e)$\carbon[13]
reaction rate is the same as that used in \citet{Piersanti2017}
(G. Mart\'{i}nez-Pinedo 2017, private communication).

The Urca process operates via a number of pairs
\citep[e.g.,][]{Tsuruta1970}.  The most prominent in CO mixtures are
\neon[21]-\fluorine[21], \sodium[23]-\neon[23],
\magnesium[25]-\sodium[25], and \sodium[25]-\neon[25]
\citep{Iben1978a}.  Throughout this work we will focus exclusively on
\sodium[23]-\neon[23], as this pair is the most abundant initially and
is also directly produced by carbon burning.  Therefore, we when refer
to any Urca-related concept (e.g., location of the Urca shell)
throughout this paper, it is always with this pair in mind. The
threshold density for this pair is
$\approx \unit[1.9\times10^9]{\gcc}$.

We use the HELM equation of state \citep{Timmes2000b}, as implemented
in \MESA, throughout the star.\footnote{When this work started, the
  Skye EOS \citep{Jermyn2021} did not yet exist and the numerical
  properties of PC \citep{Chabrier1998, Potekhin2000}, as implemented
  in \MESA, were not up to the task.  Future work should certainly
  prefer Skye to HELM in these conditions.} (This includes regions
that would normally be covered by other EOSes.) Our motivation for
this is that some of the experiments we undertake benefit from using
the identical EOS through the star as EOS blends can be a source of
spurious entropy generation.

\section{A Simple Reference Model}
\label{sec:toy}

\begin{figure}
  \includegraphics[width=\columnwidth]{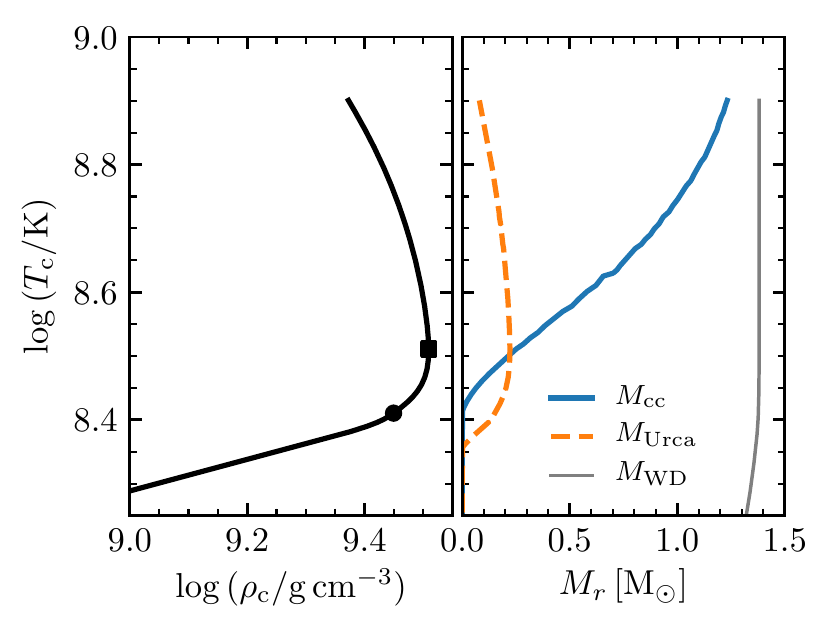}
  \caption{Evolution of our toy model.  The left panel shows the
    evolution of the central temperature and density.  Since the
    central temperature monotonically increases, it is a useful
    parameter for following the progress of the simmering.  The right
    panel shows, at the time the model has that central temperature,
    the mass coordinate of the edge of the convective core and the
    mass coordinate of the Urca shell.  The circle marks the point
    when the convective core forms; the square marks the point where
    the convective core first encompasses the Urca shell.}
  \label{fig:RhoTM}
\end{figure}

We begin with a WD model that contains 70\% \carbon[12] and 30\%
\oxygen[16] by mass.  It does not contain any Urca-pair isotopes.  We
then evolve this model in \MESA\ with composition changes from nuclear
reactions turned off (\texttt{dxdt\_nuc\_factor = 0}).  This means
that the WD remains chemically homogeneous throughout its evolution
and thereby removes the necessity to model chemical mixing.\footnote{This circumvents the issues and shortcomings related to the interplay of reactions and mixing in \MESA.}  It also
means that the outer convective boundary can be easily located via the
Schwarzschild criterion.\footnote{%
The location of the outer convective boundary is influenced by
the difference in composition between the contents of convection zone
and the unmixed and unburned outer layers of the WD \citep{Piro2008b}.}

While this may seem an extreme assumption, only
$\approx \unit[0.02]{\Msun}$ of \carbon[12] are required to be burned
to raise the WD to the temperature at which simmering ends (and the
stopping condition of our calculations) of
$\Tc = \unit[8\times10^8]{K}$ \citep{Woosley2004}.  Therefore, the
bulk composition does not dramatically change during simmering.  This
simplified model will provide a blank canvas on which to paint our
thoughts.

Figure~\ref{fig:RhoTM} summarizes the evolution of this model.  The
left panel shows the evolution of the central density and temperature.
The WD is initially compressed by the accretion, but then as carbon
burning proceeds and the convective core heats up, the WD
hydrostatically adjusts and the central density falls.  The right
panel shows the extent of the central convective zone.  The Urca shell
(defined as the place where $\recap = \rbeta)$ has a typical mass
coordinate $M_r \approx \unit[0.2]{\Msun}$ and is encompassed by the
convective core.

\subsection{Timescales}

The long timescale in the problem is the mass growth timescale of the
WD,
\begin{equation}
  t_{\dot{M}} \equiv \frac{M}{\dot{M}} \sim \unit[10^{14}]{s} \left(\frac{\dot{M}}{\unit[3\times10^{-7}]{\Msunyr}}\right)^{-1}~,
\end{equation}
assuming $M \approx \unit[1.4]{\Msun}$.  Carbon burning effectively
begins once its reaction timescale falls below this timescale and can
run away once its energy generation rate exceeds rate of cooling via
thermal neutrinos.

The short timescale is the dynamical timescale of the WD,
\begin{equation}
  t_{\rm dyn} \equiv \frac{1}{G\rho} \sim \unit[0.1]{s} \left(\frac{\rho}{\unit[10^9]{\gcc}}\right)^{-1/2}.
\end{equation}
As the timescale for carbon burning approaches the dynamical time, a
deflagration will form.

There is a local heating timescale,
$t_{\mathrm{heat, local}} = c_P T/\epsnuc$, but the convection zone is
well coupled and so local energy release is distributed across the
whole convection zone.  (The breakdown of this assumption is what
ultimately ends simmering.)  Therefore, following \citet{Piro2008a},
we use the heating timescale associated with the the convection zone
itself
\begin{equation}
  \label{eq:cooling}
  t_{\mathrm{heat}} = \left(\frac{d \ln \Tc}{d t}\right)^{-1}.
\end{equation}

Within the convection zone there is a minimum electron capture
timescale, $t_{\rm ec, min}$, and a minimum beta decay timescale,
$t_{\beta, \rm min}$.  These are located at the center and at the
outer edge of the convection zone, respectively.  However, the
reaction rates are significantly lower throughout most the convection
zone, so an average timescale is also of interest.  Define the
mass-weighted average of a quantity $x$ over the convective zone as,
\begin{equation}
  \left<x\right> \equiv \frac{1}{\Mcc} \int_0^{\Mcc} x\,dm ~.
\end{equation}
Then these average timescales are
$t_{\mathrm{ec}} = \left<\recap\right>^{-1}$ and
$t_{\beta} = \left<\rbeta\right>^{-1}$.

We can define an approximate advective mixing timescale across the
convective core as
\begin{equation}
  t_{\rm mix} = \int_0^{\rcc} \frac{dr}{v_{\rm conv}} ~,
\end{equation}
where $v_{\rm conv}$ is the convective velocity reported by \MESA\ via
mixing length theory (MLT).

\begin{figure}
  \includegraphics[width=\columnwidth]{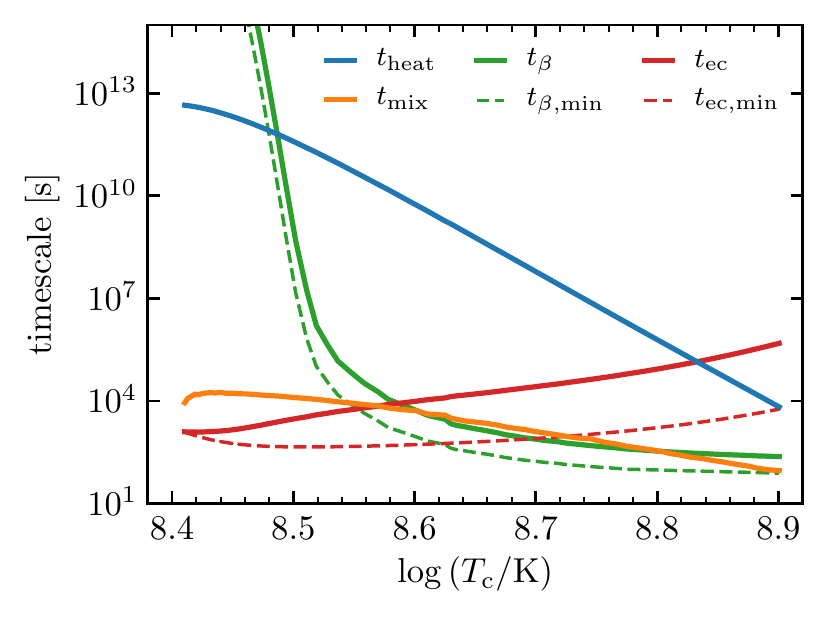}
  \caption{Timescales (as defined in the text) for our simple reference
    model.}
  \label{fig:timescales}
\end{figure}

Figure~\ref{fig:timescales} plots these timescales.  The fact that the
reaction timescales are often of order the mixing timescale is part of
what makes modeling this process challenging.  Except at the highest
central temperatures, the reaction and mixing timescales remain well
below the heating timescale.  This means it can be reasonable to think
about the interaction of reactions and mixing occurring at a fixed
structure.

\subsection{Neutrino luminosities}

% The convection is efficient, and so the temperature gradient in the
% convection zone is close to $\gradad$.

Given the sequence of WD structures from our toy reference model, we
can calculate what the luminosity in Urca process neutrinos would be
for assumed abundances and mixing properties.

Assume a total mass fraction of the Urca pair $X_{\rm U}$ that is
constant throughout the convective region.  We define the local
fraction of the parent isotope as $f_{\rm p}$.  Then the luminosity is
\begin{equation}
  L_{\rm U} = \frac{X_{\rm U}}{A_{\rm U} \amu} \int_0^{\Mcc} \left[f_{\rm p} \nuecap + (1-f_{\rm p}) \nubeta \right] dm ~,
  \label{eq:Lu}
\end{equation}
where $A_{\rm U} = 23$, the atomic mass number of our Urca pair of interest.
To estimate $f_p$, we work in the same diffusive mixing framework as is used in \MESA.  Assuming steady state abundances, 
\begin{equation}
  -f_{\rm p} \recap + (1-f_p) \rbeta + \ddm{}\left(\sigma \ddm{f_{\rm p}}\right) = 0
  \label{eq:ssmix}
\end{equation}
where $\sigma = D(4\pi r^2\rho)^2$ is the Lagrangian diffusion
coefficient.  With the knowledge of $D$, we can then solve this
equation to obtain the profiles within the convection zone.

\subsubsection{No mixing ($ D = 0$)}
For $D = 0$, the problem is the same as for Urca cooling that occurs
in stable radiative regions (i.e., thermal Urca process).  This is
effectively the minimum amount of cooling that can occur. The
abundances are set by the local rate balance condition
$f_{\rm p} \recap = (1-f_{\rm p}) \rbeta$.  As a result,
\begin{equation}
  L_{\mathrm{U}, 0} = \frac{X_{\rm U}}{A_{\rm U} \amu} \int_0^{\Mcc}
  \left(\frac{\rbeta \nuecap + \recap \nubeta}{\recap + \rbeta}\right) dm ~,
  \label{eq:Lu0}
\end{equation}
where $\nuecap$ and $\nubeta$ are respectively the specific neutrino
loss rates from electron capture and beta decay.

\subsubsection{Instantaneous mixing $(D = \infty)$}

We can also make the extreme assumption that the convection zone is
instantaneously mixed.  This will maximize the rate at which the
convective Urca process can operate.  Within the mixed region, uniform
abundances of the parent/daughter Urca isotopes are set by the balance
of the mass-weighted rates through the convection zone.  Then the
total Urca-process neutrino luminosity from the convective zone is
\begin{equation}
  L_{\mathrm{U}, \infty} = \frac{X_{\rm U} \Mcc}{A_{\rm U} \amu} \left(\frac{\left<\rbeta\right> \left<\nuecap\right> + \left<\recap\right> \left<\nubeta\ \right>}{\left<\recap\right> + \left<\rbeta\right>}\right)~.
  \label{eq:Luinf}
\end{equation}

\subsubsection{Results}

\begin{figure}
  \includegraphics[width=\columnwidth]{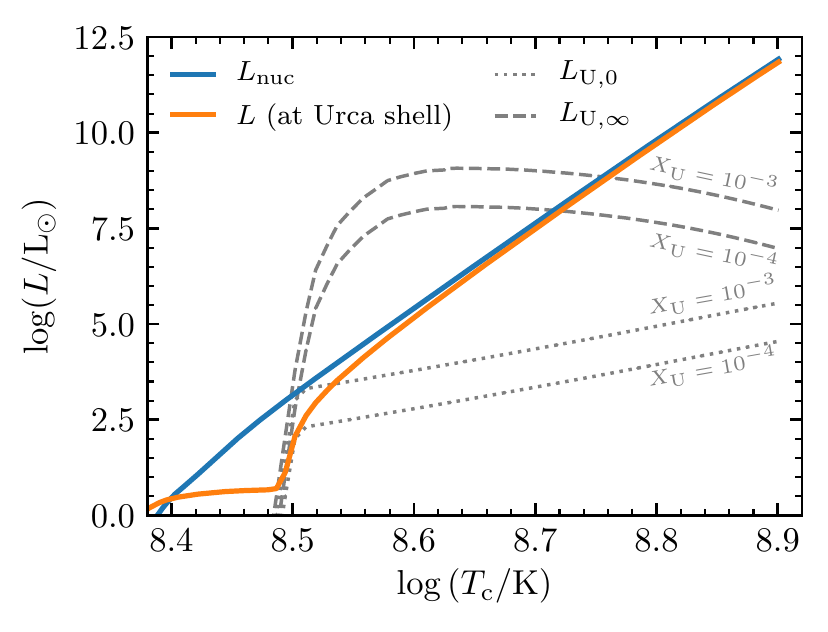}
  \caption{Characteristic luminosities as a function of central temperature.  Solid lines show the total nuclear luminosity of the \MESA\ models and the local luminosity at the Urca shell.  Dashed lines are estimates of the minimum Urca-process neutrino luminosities (Equation~\ref{eq:Lu0}). Dotted lines are estimates of the maximum Urca-process neutrino luminosities (Equation~\ref{eq:Luinf}).}
  \label{fig:Lus}
\end{figure}

Figure~\ref{fig:Lus} puts the luminosity bounds from our estimates in
Equations~\eqref{eq:Lu0} and \eqref{eq:Luinf} in context.  As a
function of model central temperature, these are compared to the total
nuclear luminosity.  The convection zone encompasses the Urca shell
around $\logTc \approx 8.5$.  From then, until $\logTc \gtrsim 8.75$,
the nuclear luminosity from carbon burning is well below the maximum
neutrino luminosity, meaning the Urca process has the opportunity for
maximum impact.  Above that temperature, the carbon burning is sufficiently rapid that it cannot significantly influence the global energetics. 

\begin{figure}
  \includegraphics[width=\columnwidth]{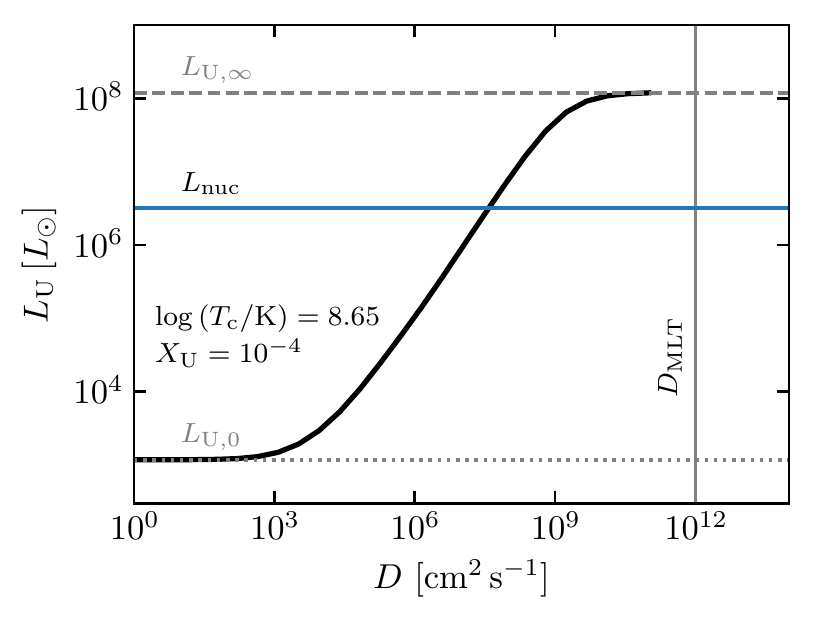}
  \caption{Neutrino luminosities from evaluating
    Equation~\eqref{eq:Lu} for the steady-state abundance profiles
    associated with a range of constant diffusion coefficients.  This
    assumes the stellar structure of the \MESA\ model at the indicated
    point in the evolution.}
  \label{fig:constantD}
\end{figure}

In Figure~\ref{fig:constantD}, we show the result of numerically solving
Equation~\eqref{eq:ssmix}, assuming the structure from the \MESA\
model at a particular time, for a range of constant $D$.  This
confirms the limiting behaviors derived analytically.  It also
indicates that the value of $D$ reported by \MESA\ MLT puts us in the
limit of $L_{\rm U} \approx L_{{\rm U}, \infty} \gg L_{\rm nuc}$.

% mesa MLT seems to give 1e12 or so for D
% $L \sim 4 \pi r^2 \rho v_c^3$, implying a diffusivity $D \sim v_c H$.

\subsection{Extent of convective core}

% The previous subsection suggests the strong 

If we assume that there is a central region that is being mixed
infinitely fast, then the neutrino luminosity also depends on where
the outer edge of this mixed region is placed.  The estimate of
Equation~\eqref{eq:Luinf} assumes this mixed region corresponds to the
convective region reported in the \MESA\ models ($M_{\rm cc}$).

\begin{figure}
  \includegraphics[width=\columnwidth]{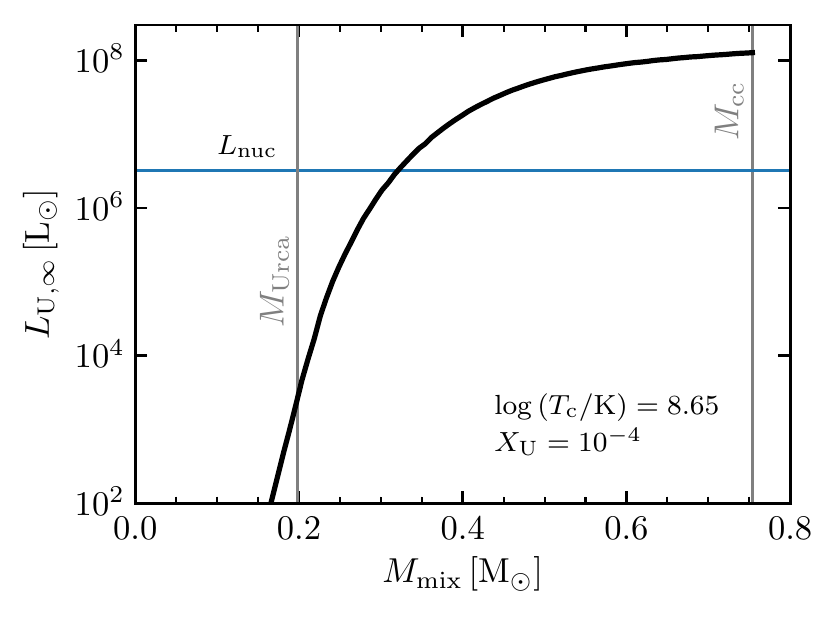}
  \caption{Illustration of finding $M_{\rm mix, U}$, the point when $L_{\rm U, \infty} = L_{\rm nuc}$.  This will be intermediate between the location of the Urca shell ($M_{\rm Urca}$) and the unmodified location of the convective core ($M_{\rm cc}$). \label{fig:finding_MmixU}}
\end{figure}

\begin{figure}
  \includegraphics[width=\columnwidth]{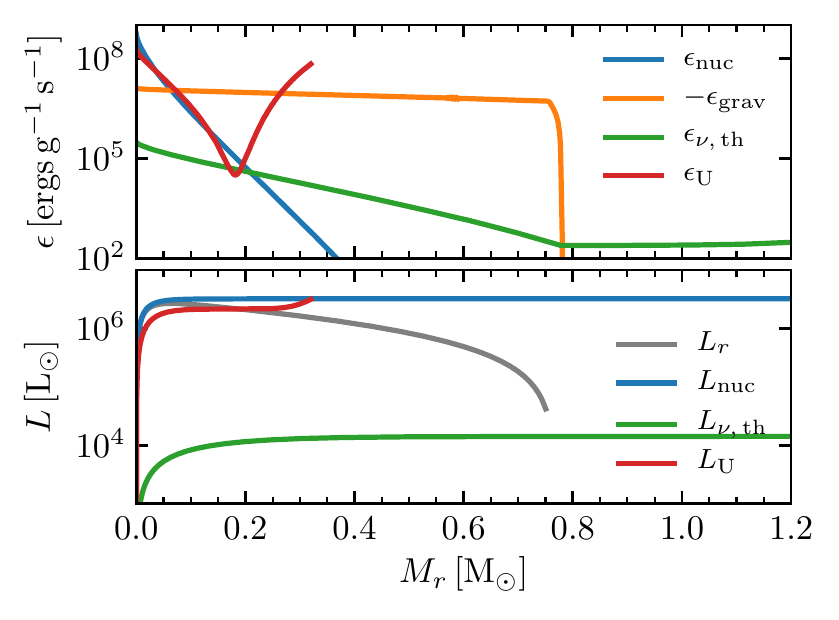}
  \caption{Local energy generation rates (top panel) and integrated
    luminosities (bottom) panel for the model shown in
    Figure~\ref{fig:finding_MmixU}. \label{fig:eps_and_L}}
\end{figure}

To illustrate the potential effect of the Urca process, we evaluate the
neutrino luminosity assuming a centrally mixed region of size
$M_{\rm mix} \le M_{\rm cc}$.  We define a critical extent
$M_{\rm mix, U}$ that occurs when $L_{\rm U, \infty} = L_{\rm nuc}$.
Figure~\ref{fig:finding_MmixU} visually illustrates this process for a
single model snapshot.  This calculation is not self-consistent as the
temperature-density structure of the star is held fixed independent of
size the hypothetically mixed region.  Figure~\ref{fig:eps_and_L}
shows the energy generation rate and cumulative luminosity associated
with the Urca process superimposed on the background model.  This
illustrates that in a steady state, the neutrino cooling happens near
the edges of the convection zone.  In fact, the Urca shell is the
location of the minimum neutrino cooling rate.

\begin{figure}
  \includegraphics[width=\columnwidth]{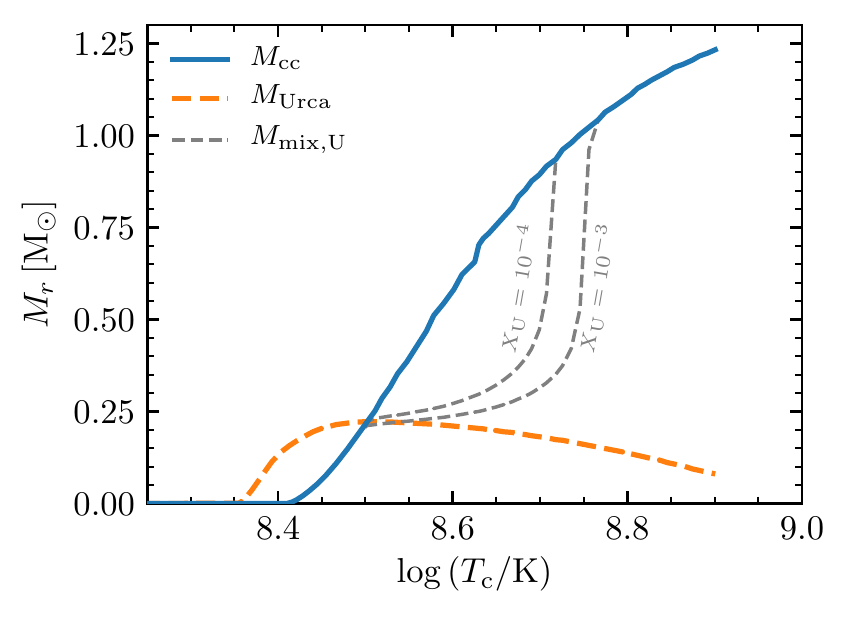}
    \caption{Size of the mixed region necessary to balance nuclear
    luminosity.  Grey dashed lines show this extent for the two
    indicated values of the Urca isotope mass fraction.  Balance is
    only possible when the grey curves are visible. \label{fig:MmixU}}
\end{figure}

By repeating the procedure described above on the sequence of stellar
structures from the simplified model, we map out how the size of this
mixed region changes with central temperature.  Figure~\ref{fig:MmixU}
shows that the mixed region must generally extend somewhat beyond the
Urca shell, but that the region may be significantly restricted.

This likely remains an overestimate of the size of the core.  The
luminosity profile throughout the convective region is affected by the
fact that it is not in thermal equilibrium as work to expand the star
is being performed on the heating timescale \citep{Piro2008b}.  If the
Urca cooling were to more nearly balance the nuclear heating, then the net
heating (and expansion) timescale would lengthen and expansion work
would be done less rapidly.  In the extreme limit where the heating
and cooling balanced, the evolution would instead proceed on the
timescale of the composition change.

% Another relevant effect may be that the growth of the convection
% zone dilutes the mass fraction of the Urca isotope.  Core grows a
% little bit, Urca falls, has to re-equilibrate.

\section{Onset of Convective Carbon Burning}

When central carbon burning begins, the star is relatively quickly
destabilized.  The region where the carbon fusion is occurring is
almost always above the \sodium[23] threshold density, so that the
$\sodium[23](e^-, \nu_{e})\neon[23]$ reaction almost always occurs.
This provides more nuclear energy per carbon burnt and also means that
the Urca-pair composition of the convection zone will be primarily the
\neon[23].  Eventually, the convection zone grows to the location of
the Urca shell. At that time there will be a \neon[23] abundance that
is the sum of the initial \sodium[23] abundance and the \sodium[23]
produced thus far from carbon burning.  This means there is a
threshold metallicity below which the Urca-pair abundance in the
convection zone at the time of formation of the Urca shell is
approximately constant.

\section{The Convection Zone Reaches the Urca Shell}

There is a maximum luminosity that can be carried (radiatively) by
stably stratified material, corresponding to the point where
$N^2 = 0$.  Typically, this corresponds to the adiabatic
stratification ($\gradT = \gradad$).  However, there can be an
additional enhancement in the presence of stabilizing composition
gradients.  That is, we have $\gradad < \gradT \le \gradad + B$, and
so a luminosity
\begin{equation}
  L_{\rm max} = \frac{16 \pi a c G M_r T^4}{3 \kappa P} (\gradad + B)~.
\end{equation}
When the composition dependence of the pressure is through the
electron fraction, as in degenerate material, the Brunt term for
stability can be written as
\begin{equation}
  \label{eq:17}
  B = -\frac{1}{\chi_T}
      \left(\frac{\partial \ln P}{\partial \ln \Ye}\right)_{\rho,T}
      \frac{d \ln \Ye}{d \ln P}~.
\end{equation}

We can make a simple estimate for the value of $B$ in these
circumstances.  For a cold plasma with degenerate electrons and ideal
ions,
$\left(\partial \ln P / \partial \ln \Ye\right)_{\rho,T} \approx 4/3$
and $\chi_T \approx 4kT/(\bar{Z} E_\mathrm{F})$.  Given a mass
fraction $X_{\rm U} \ll 1$ of an Urca isotope with mass number
$A_{\rm U}$, the change in electron fraction associated with the
reaction is $\Delta \Ye = -X_{\rm U}/A_{\rm U}$.  We can rewrite

\begin{equation}
  \frac{d \ln \Ye}{d \ln P} = \frac{1}{\Ye}\frac{d \Ye}{dr} \frac{dr}{d\ln P} \approx \frac{1}{\Ye}\frac{\
\Delta \Ye}{\Delta r} H_{P}~,
\end{equation}
where $\Delta r$ reflects the size of the Urca shell.  The equilibrium
shifts when the chemical potential changes by $\approx kT$, so
$\Delta r \approx 4 (kT/\EF) H_{P}$.  For the typical conditions
associated with \sodium[23] and carbon simmering, this implies
\begin{align}
  B_{\rm max} & \approx X_{\rm U} \left(\frac{\Zbar/A_{\rm U}}{\Ye}\right) \left(\frac{4 kT}{\EF}\right)^{-2} \\ & \sim 0.6 \left(\frac{X_{\rm U}}{10^{-4}}\right) \left(\frac{T}{\unit[2\times10^8]{K}}\right)^{-2}~.
\end{align}
This suggests that the composition gradient can provide some
additional stabilization but not so much that convection would be
unable to penetrate at the radius around the threshold density and set
up the Urca shell.

\section{Summary and Conclusions}

By using toy models and sacrificing some aspects of self-consistency,
I made estimates about neutrino luminosities and convective core sizes
that are relevant to the operation of the convective Urca process
(Section~\ref{sec:toy}).

My initial goal was to produce \MESA\ models of white dwarfs
undergoing high-density convective carbon burning and exhibiting local
energy conservation.  That seems a key milestone before one can hope
to characterize the effect of a convective region containing an Urca
shell in stellar models. (Subsequently, one can start to tackle
questions of whether what is happening in the 1D stellar models is an
accurate reflection of what happens in a real star.)  Despite
significant effort, I failed to realize that goal, but the attempts
resulted in a number of improvements to \MESA.  I'm still not sure if
it is possible, but it is less impossible than when I started.

% \begin{acknowledgments}

\vspace{2em}

  I thank Carles Badenes, Lars Bildsten, Adam Jermyn, Samuel Jones,
  Stephen Justham, Daniel Lecoanet, Gabriel Mart{\'{\i}}nez-Pinedo,
  H\'{e}ctor Mart{\'{\i}}nez-Rodr{\'{\i}}guez, Anthony Piro, Philipp
  Podsiadlowski, Eliot Quataert, Donald Willcox, Stan Woosley, and
  Michael Zingale for helpful conversations on this and related topics.

Josiah was supported by NASA through Hubble Fellowship grant
HST-HF2-51382.001-A awarded by the Space Telescope Science Institute,
which is operated by the Association of Universities for Research in
Astronomy, Inc., for NASA, under contract NAS5-26555, by the
A.F. Morrison Fellowship in Lick Observatory, by the NSF through grant
ACI-1663688, via support for program number HST-GO-15864.005-A
provided through a grant from the STScI under NASA contract
NAS5-26555, and by his wife Annelise Beck.
Josiah's computer was supported by Bill Paxton's Stanford
yearbooks and an Interlisp reference manual.

%\end{acknowledgments}

\clearpage

\appendix

\section{Conservation Laws and Thermodynamics}
\label{sec:laws}

Understanding the way that energy conservation is encoded in the
stellar structure equations can prove confusing, especially in the
presence of composition changes due to nuclear reactions and chemical
transport processes.  Here we carefully work through how conservation
principles and thermodynamics lead to the stellar structure equations.
We then discuss various limitations and approximations that are (or have been)
present in the \MESA\ implementation of these equations.

% A star is a system with properties that are functions of both space
% and time.  The classic thermodynamic systems one envisions have
% homogeneous properties (i.e. the system has \textit{a} temperature)
% and in equilibrium thermodynamics there is no time evolution.  Clearly
% modeling a star requires additional effort.

% Thinking in the continuum limit is essentially equivalent into
% dividing our star into many infinitesimal systems (but not so
% infinitesimal that thermodynamics breaks down).  How do we think of
% each of these individual systems?  Given the fundamental role of
% energy transport in stars, these are clearly not isolated systems.
% Sometimes, we think of them as ``closed'' systems, where there can be
% an exchange of heat with their surroundings.  In reality, they are
% ``open'' systems that exchange both heat and matter with their
% surroundings.  The fact that the systems can exchange material is
% obviously of critical importance in understanding the global evolution
% of the star (e.g., how much nuclear fuel can be burned).  But we also
% want to understand when it is important in the local thermodynamic
% sense.

\subsection{Baryon Number}

Since nuclear reactions involve changes in mass, it is important that we
work in the conserved physical quantity, which is the baryon number
$\NB$.  Denoting Avogadro's number by $\NA$, the atomic mass unit is
$\amu = 1\,{\rm g}/ \NA$. The number abundance of every species is
defined with reference to the baryonic number density,
$Y_i \equiv n_i/\nB$.  The total baryonic mass density is
$\rho = \nB / \NA$, so that $1/\rho$ is the specific volume.  Local
charge neutrality and local thermodynamic equilibrium (LTE) determine
a unique solution for the ionization state of each isotope.  Thus, the
composition is completely specified by a set of number abundances
$\{Y_i\}$ for all nuclear isotopes and electrons are not a separately
tracked species.  The mass fractions are $X_i = A_i Y_i$, where $A_i$
is the nucleon number; by construction $\sum_i X_i = 1$.

The specific rest mass energy density is
\begin{equation}
  e_{\rm rest} = \sum_i \NA M_i c^2 Y_i
  \label{eq:rest-mass}
\end{equation}
Our definition of rest mass is the atomic mass (or atomic mass
excess).  The atomic mass includes the rest mass of the nucleus and
electrons; it also includes the electron binding energy, which should
not be present when material is fully ionized, but is typically
negligible.

% Note that since the masses are

% \begin{equation}
%   M = \sum_i m_i   
% \end{equation}

% Therefore, we ought to be careful in defining our Lagrangian
% coordinate in terms of baryon number, rather than mass.

\subsection{Species and Mass Conservation}

The baryonic mass density of an individual species
($\rho_i = \rho X_i$) is not conserved, but rather obeys the
advection-diffusion equation,
\begin{equation}
  \ddt{(\rho X_i)} + \grad \vecdot \left(\rho X_i \vec{v}\right) =
  \grad \cdot \left(D \rho \grad X_i \right) + \rho \dot{X}_i ~,
  \label{eq:one-species}
\end{equation}
where $D$ is a (species-independent) diffusion coefficient and
$\dot{X}_i = A_i \dot{Y}_i$ represents the rate of change of species
$i$ due to nuclear reactions.  Summing the individual equations yields
conservation of baryonic mass
\begin{equation}
  \ddt{\rho} + \grad \vecdot \left(\rho \vec{v}\right) = 0 ~.
  \label{eq:all-species}
\end{equation}
since
$\sum_i \grad \cdot \left(D \rho \grad X_i \right) = \grad \cdot \left(D \rho \grad \sum_i X_i \right) = 0$
and we assume the reactions conserve baryon number.

Defining the usual Lagrangian derivative,
\begin{equation}
  \DDt{} = \ddt{} + \vec{v} \cdot \grad ~,
\end{equation}
and with the help of Equation~\eqref{eq:all-species}, we can rewrite Equation~\eqref{eq:one-species} as
\begin{equation}
    \rho \DDt{X_i} = \grad \cdot \left(D \rho \grad X_i \right) + \rho \dot{X}_i~.
\end{equation}
Specializing to spherical coordinates, recall
\begin{equation}
    \ddr{} = 4 \pi r^2 \rho \ddm{}
\end{equation}
and thus we can write
\begin{equation}
    \DDt{X_i} = \ddm{} \left( D (4 \pi r^2 \rho)^2 \ddm{X_i} \right) + \dot{X}_i~.
\end{equation}
After defining the Lagrangian diffusion coefficient
$\sigma = D(4\pi r^2\rho)^2$, we arrive at the time evolution
equations that \MESA\ solves for the mass fractions $X_i$ (see
Equations 13-14 in \mesaone),
\begin{equation}
  \label{eq:local-mass-fraction}
  % \ddt{X_i}
\DDt{X_i} = \underbrace{\dot{X_i}\vphantom{\ddm{}\left(\sigma \ddm{X_i}\right)}}_{\mathrm{burn}} + \underbrace{\ddm{}\left(\sigma \ddm{X_i}\right)}_{\mathrm{mix}}.
\end{equation}
Traditionally we name the first term the ``burn'' part (the bulk term
due to nuclear reactions, where $\dot{X}_i$ comes from the
\texttt{net} module ) and the second term the ``mix'' part (the
boundary term due to convective mixing, where $\sigma$ comes from MLT).

\subsection{Thermodynamics}

The first law of thermodynamics is a statement about energy
conservation.  In \mesafour, we derived a total energy equation by
combining the first law with the momentum equation.  Chapter II of
\citet{dGM} demonstrates that the first law comes from combining the
more fundamental hydrodynamical equations that encode the conservation
of mass, momentum, and total energy (see also Appendix A in \citealt{Bauer2020}).
Assuming that there is not
element diffusion (i.e., all species move with the same bulk
velocity), their Equation~(II.39) reduces to the familiar form
\begin{equation}
  \label{eq:first-law}
  \DDt{e} = \DDt{q} - P\DDt{}\left(\frac{1}{\rho}\right)~,
\end{equation}
where $e$ is the internal energy.
% (In our \neon[22] notes, Evan and I
% demonstrate that this equation is effectively the same even when each
% species has its own diffusion velocity, so long as diffusion does not
% lead to a net mass flux.)

The ``heat flow'' term in Equation~\eqref{eq:first-law} is
\begin{equation}
  \DDt{q} \equiv \dot{q} - \frac 1 \rho \grad \vecdot \vec{J}_q ~.
  \label{eq:heat-flow}
\end{equation}
This represents the total rate of the flow of energy into a Lagrangian
parcel.  The quantity $\dot{q}$ is the specific volumetric source
term, reflecting the rate of processes that can directly remove energy
from the system (i.e., optically-thin cooling).  The quantity
$\vec{J}_q$ is the heat flow and so this term reflects energy moving
through the boundaries of a volume into adjacent regions.

% This derivation has treated the total specific energy as the sum of
% the internal, kinetic, and potential components.  This expression
% doesn't include rest mass.  (How would that enter?)

% The entropy is a quantity that can be defined in terms of the
% macroscopic characteristics of the system.  If we write $S$ in terms
% of the independent thermodynamic basis variables $(E,V,N_i)$, then
% expanding it yields the thermodynamic identity
% \begin{equation}
%   \label{eq:thermo-id}
%   T \dif S = \dif E + P \dif V - {\sum_i \mu_i \dif N_i}~,
% \end{equation}
% where $S$ is the entropy, and $T$ is the temperature.  The sum runs
% over all species present, and $\mu_i$ is the chemical potential for
% species $i$.  The specific (i.e.,~per unit mass) form of
% Equation~\eqref{eq:thermo-id} is then given by multiplying by the
% invariant $N_{\rm A}/N_{\rm B}$ to find
% \begin{equation}
%   \label{eq:specific-id}
%   \dif e + P \dif \left( \frac 1 \rho \right)
%   = T \dif s  + {\sum_i \NA \mu_i\, \dif Y_i}~.
% \end{equation}

% \citet{dGM} refer to this as the Gibbs relation.
% Thorne and Blandford call this ``first law of thermodynamics''.

\subsection{Local Energy Equation}

% To understand the energetic role of nuclear reactions, we can
% consider the star a set of locally closed systems, since the
% composition changes occur from within.  This means we can directly
% follow the arguments in MESA IV.

In \mesafour, we stated that in the stellar structure equations,
energy conservation is typically formulated by considering the energy
flow in and out of a fluid parcel.  This idea is an expression of the
physics encoded by Equation~\eqref{eq:heat-flow}.  Traditionally, we
denote the specific volumetric source term as $\epsilon$.
%
% The only mechanism by which energy can directly leave the system is
% through neutrinos and so $\dot{q} \equiv \epsilon = \epsnu$.

As discussed previously, the conserved quantity is baryon number, so
we define our Lagrangian coordinate to be the enclosed baryonic mass,
i.e., the number of enclosed baryons times the atomic mass unit.  This
means that there can be a flow of rest mass energy through the
boundary of a parcel. Then the total energy flow is
\begin{equation}
  \vec{J}_{q} = \boldsymbol{\mathcal{F}} - \sum_i D \grad \left(M_i c^2 \NA \frac{\rho X_i}{A_i}\right)~,
\end{equation}
where we represent the familiar flux of energy carried by radiation,
convection, and conduction with $\boldsymbol{\mathcal{F}}$.  Recall that 
by the definition of $L$, under spherical symmetry
\begin{equation}
  \ddm{L} = \frac{1}{\rho} \grad \cdot \boldsymbol{\mathcal{F}}.
\end{equation}
Thus taking the divergence
% \begin{equation}
%   \frac{1}{\rho} \grad \cdot \vec{J}_{q} = \ddm{L} - \sum_i \NA \frac{M_i c ^2}{A_i} \ddm{}\left(\sigma \ddm{X_i}\right)~.
% \end{equation}
and inserting into Equation~\eqref{eq:heat-flow}, we see that the
parcel obeys
\begin{equation}
  \label{eq:local-energy}
  \DDt{q} = \epsilon - \ddm{L}
  + \sum_i \NA \frac{M_i c ^2}{A_i} \ddm{}\left(\sigma \ddm{X_i}\right)~.
\end{equation}
The familiar form (Equation 54 in \mesafour) would be restored if we
neglected the rest mass.

% In a radiative zone, the diffusion coefficients are the microphysical
% ones; in a convective zone, the diffusion coefficients are replaced by
% turbulent ones.

% We will work relativitically, meaning that heating from nuclear
% reactions will be accounted for via a change in rest mass.

% In this Lagrangian picture, to understand how the energy of a
% fluid parcel is changing, we account for the specific (i.e.,~per unit
% mass) rate of energy injection into the parcel, $\epsilon$, and the
% specific rate of energy flow through the boundaries
% ($\partial L/\partial m$; $L(m)$ is the luminosity profile and $m$ the
% Lagrangian mass coordinate).

% Now what about this term.
% \begin{equation}
%   {\sum_i \frac{N_{\mathrm{A}}}{N_{\mathrm{B}}} \mu_i \ddm{F_i}}
% \end{equation}
% This term is not in MESA.
% If we're diffusive, then the flux is $F_i = \sigma \ddm{N_i}$, so
% \begin{equation}
%   \label{eq:3}
%   {\sum_i \frac{\NA}{A_i} \mu_i \ddm{}\left(\sigma(m) \ddm{X_i}\right)}
% \end{equation}
% where we have used the fact that $Y_i = X_i / A_i$.

\subsection{Nuclear Reactions}

Combining Equations~\eqref{eq:first-law} and
\eqref{eq:local-energy} gives
\begin{equation}
  \DDt{e} - \frac{P}{\rho^2}\DDt{\rho} = \epsilon - \ddm{L}
  + \sum_i \NA \frac{M_i c ^2}{A_i} \ddm{}\left(\sigma \ddm{X_i}\right)~.
\end{equation}
We separate out the rest mass energy density,
\begin{equation}
  \DDt{e} = \DDt{e_{\rm rest}} + \DDt{e_{\rm int}} = \sum_i \NA M_i c^2 \DDt{Y_i} + \DDt{e_{\rm int}}
\end{equation}

Taking Equation~\eqref{eq:local-mass-fraction} and dividing by
$A_i$ gives an equation for $\Dif {Y_i}/\Dif t$. Inserting this gives
\begin{equation}
  \label{eq:combined}
  \DDt{e} + \sum_i \NA M_i c^2 \dot{Y}_i
  - \frac{P}{\rho^2}\DDt{\rho}
  = \epsilon - \ddm{L}.
  % \underbrace{N_i \lambda_i}_{\mathrm{burn}} - \underbrace{\ddm{F_i}}_{\mathrm{mix}}
\end{equation}
where the rest mass flow term has been canceled by the mixing-related
part of the $\Dif{e}/\Dif t$ term.  A typical definition of $\epsnuc$
\citep[e.g., Equation 11 in][]{Hix2006} is
\begin{equation}
  \epsnuc = -\sum_i \NA M_i c^2 \dot{Y_i} ~,
  \label{eq:epsnuc-hix}
\end{equation}
and we immediately see that term is present in
Equation~\eqref{eq:combined}.  If rest mass is included in the
definition of energy, then $\epsnuc$ is not separately inserted into
$\epsilon$ in Equation~(\ref{eq:local-energy}).  The value of
$\epsilon$ is only optically-thin cooling, which consists of both
thermal ($\epsnuth$) and nuclear ($\epsnunuc$)
neutrinos.  The specific nuclear neutrino loss rate is
\begin{equation}
  \epsnunuc = \sum_i \langle E_\nu \rangle \dot{Y_i}~.
  \label{eq:epsnu-nuc}
\end{equation}
So we now have
\begin{equation}
    \DDt{e} 
  - \frac{P}{\rho^2}\DDt{\rho}
  = \epsnuc - \epsnunuc - \epsnuth
  - \ddm{L} ~.
  \label{eq:final}
\end{equation}
This is the usual stellar structure equation.

\subsection{\MESA\ Source Terms}

Section 3.2 in \citet{Paxton2019} describes the current definition of
\epsnuc\ in \MESA.  Importantly, in contrast to the definition given
in Equation~\eqref{eq:epsnuc-hix}, \MESA\ already subtracts off the
neutrinos in its definition, so the variable \texttt{eps\_nuc}
(provided by the \texttt{net} module) is equivalent to
$\epsnuc - \epsnunuc$.  The variable \texttt{non\_nuc\_neu}
(provided by the \texttt{neu} module) is equivalent to
$\epsnuth$.

The original \texttt{weaklib} treatment of weak reactions in \MESA\
\citep{Paxton2011} included the chemical potential of the electron in
$\epsnuc$.  This term conceptually belongs in $\epsgrav$, but it was
easier to include in $\epsnuc$ because you can evaluate that
contribution as you sum over the reaction rates.  One problem with
that approach is that it then only includes composition changes due to
nuclear reactions.  When element diffusion or convective mixing is
creating a rate of change in the composition comparable to (or greater than) nuclear
reactions and this is occurring in a region where kinetic chemical
potentials are important, then \MESA\ will then give the wrong answer.
This is a critical shortcoming in the \MESA\ calculations of the
simmering phase presented in \citet{MartinezRodriguez2016} and is
discussed from a different perspective in \citet{Schwab2017a}.

Significant software development effort has gone into correcting this
issue and the details are expected to be discussed in a forthcoming
\MESA\ instrument paper.  In short, the relevant composition-related
term can be calculated from a specialized finite difference of the
internal energy and included (by default) in $\epsgrav$.  One caveat
remains that the thermodynamic basis, $(\rho, T, \{X_i\})$, involves
the abundances of all isotopes.  However, in many cases the EOS basis
is not the full composition basis because some component EOSes in \MESA\ 
reduce the composition basis from all species to a few numbers (e.g.,
$X,Y,Z$ or $\bar{A},\bar{Z}$).  This can limit the ability to calculate the physically relevant energy
changes.  For example, a change in the metal abundance pattern that
left $Z$ unchanged would be invisible to the EOS.  The fact that the
new Skye EOS \citep{Jermyn2021} uses all isotopes with significant
mass fractions in its basis means this caveat should not pose a practical concern
for future work on this problem using \MESA.

\clearpage

\section{An Astrophysical Apparition: The Papaloizou-Pringle Patronus}
\label{sec:deathotter}

While in graduate school at Berkeley, I was working on modifying the
ZEUS-MP2 code \citep{Hayes2006} to include an $\alpha$-viscosity
treatment in preparation for work evolving white dwarf merger remnants
\citep{Schwab2012}.

I ran many test problems that followed the evolution of the
equilibrium torus solutions considered by \citet{Papaloizou1984}.
During one such experiment, my analysis scripts produced the following
contour plot of density.  The image has been manipulated only by
rotation (such that the equatorial plane is now vertical) and cropping
(to focus attention on the relevant feature).

This figure shows the appearance of a powerful and mysterious otter,
who is perhaps blowing the fiery bubble that will give rise to our
universe.  Future work should explore whether spectral methods
\citep[e.g.,][]{Burns2020} are more likely to produce such
illuminating results.

\vspace{1cm}
\begin{center}
   \includegraphics[]{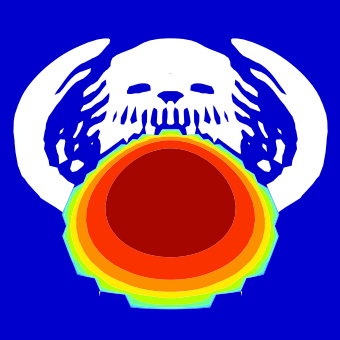}
\end{center}

\clearpage

\bibliography{paper.bib}

\end{document}